\documentclass[12pt]{article}
\usepackage{amsmath}
\def\a{\alpha}
\def\b{\beta}

\def\d{\delta}
\def\e{\epsilon}

\def\g{\gamma}
\def\p{\psi}

\def\r{\rho}
\def\s{\sigma}

\def\vf{\varphi}

\def\be{\begin{equation}}
\def\ee{\end{equation}}
\def\arr{\begin{array}{rll}}
\def\ea{\end{array}}
\def\bea{\begin{eqnarray}}
\def\eea{\end{eqnarray}}

\def\N2{$N{=}2$}

\def\>{\rangle}
\def\<{\langle}
\def\+{\dagger}
\def\={\ =\ }

\begin{document}
\renewcommand{\thefootnote}{\fnsymbol{footnote}}
\begin{titlepage}
\setcounter{page}{0}
\vskip 1cm
\begin{center}
{\LARGE\bf  $N=4$ superconformal mechanics}\\
\vskip 0.5cm
{\LARGE\bf from the $su(2)$ perspective}\\
\vskip 2cm
$
\textrm{\Large Anton Galajinsky\ }
$
\vskip 0.7cm
{\it
Laboratory of Mathematical Physics, Tomsk Polytechnic University, \\
634050 Tomsk, Lenin Ave. 30, Russian Federation} \\
{Email: galajin@tpu.edu.ru}
\vskip 0.5cm

\end{center}
\vskip 1cm
\begin{abstract} \noindent
The issue of constructing an $N=4$ superconformal mechanics in one dimension is reconsidered with a special emphasis put on the realizations of the $su(2)$--subalgebra in the full $su(1,1|2)$--superalgebra.
New dynamical realizations of $su(1,1|2)$ are constructed which describe an interaction of the $(0,4,4)$--supermultiplet with either the $(1,4,3)$--, or $(3,4,1)$--, or $(4,4,0)$--supermultiplet.
A relation of the $N=4$ superconformal mechanics with massive superparticles propagating near the black hole horizons is discussed. Background geometry associated with the model based on the $(4,4,0)$--supermultiplet is identified with the near horizon limit of the $d=5$, $N=2$ supergravity interacting with one vector multiplet.

\end{abstract}

\vspace{0.5cm}

\end{titlepage}

\renewcommand{\thefootnote}{\arabic{footnote}}
\setcounter{footnote}0

\noindent
{\bf 1. Introduction}\\

There are several reasons to be concerned about an $N=4$ superconformal mechanics in one dimension. One the one hand, the $N$--extended $d=1$ supersymmetry has
a number of peculiar features which are strikingly distinct from higher dimensional analogues. In particular, some of the $N=4$, $d=1$ supermultiplets cannot be obtained by dimensional reduction
from $d>1$ off--shell supermultiplets. Constraints describing on--shell supermultiplets in higher dimensions may lead to $d=1$ off--shell supermultiplets after a dimensional reduction. The number of physical bosonic degrees of freedom does not have to match the number of physical fermions.

On the other hand, it was conjectured in \cite{claus,gibb} that the study of superconformal models in $d{=}1$ might provide a new insight into a microscopic description of extremal black holes. In particular, this proposal stimulated extensive resent studies of the
$N=4$ superconformal Calogero models \cite{W}--\cite{IFL1}.

Another line of research on the superconformal mechanics motivated by the work in \cite{claus,gibb} concerns the construction of superconformal particles propagating on the near horizon extremal black hole backgrounds (see e.g. \cite{BIK3}--\cite{BelK} and references therein).
It turns out that such systems can be viewed as the conventional superconformal mechanics written in another coordinate basis \cite{BIK3}--\cite{gn}. It is believed that these models will help to better understand a precise relation between the supergravity Killing spinors and the supersymmetry charges of the superparticles propagating on the curved backgrounds.

It should be mentioned that the superalgebra $su(1,1|2)$ which we discuss in this work is a particular instance of the most general $N=4$, $d=1$ superconformal algebra corresponding to the exceptional one--parameter supergroup $D(2,1;\alpha)$. Various dynamical realizations of $D(2,1;\alpha)$ have been recently studied in \cite{ikl}--\cite{TO}.
As far as the proposals in
\cite{claus,gibb} are concerned, the status of the $D(2,1;\alpha)$--superconformal mechanics is still to be better understood.

There are several competing approaches to the construction of a superconformal mechanics. These include the superfield approach, the method of nonlinear realizations, and the direct construction of a representation of the
$su(1,1|2)$--superalgebra within the Hamiltonian framework. While the superfield formulation seems to be the most powerful means, the Hamiltonian approach automatically yields
an on--shell component formulation which is free from non--dynamical auxiliary fields. In some cases it also offers notable technical simplifications in describing interacting $N=4$, $d=1$ supermultiplets \cite{glp,glp1}.

The goal of this work is to reconsider the issue of constructing  an $N=4$ superconformal mechanics in one dimension with a special emphasis put on the possible dynamical realizations of the $su(2)$--subalgebra in the full $su(1,1|2)$--superalgebra. In Sect. 2 we demonstrate that any representation of $su(2)$ in terms of phase space functions can be automatically extended to a representation of the full $su(1,1|2)$--superalgebra. In Sect. 3 we discuss various examples which yield the $N=4$, $d=1$ supermultiplets of the type $(1,4,3)$, $(3,4,1)$, and $(4,4,0)$ as well as construct interacting systems which describe couplings of the former supermultiplets to ($n$ copies) of the $(0,4,4)$--supermultiplet. In Sect. 4 we discuss a link to superparticles propagating near the black hole horizons and identify a curved background associated with the supermultiplet of the type $(4,4,0)$ with the near horizon limit of the $d=5$, $N=2$ supergravity interacting with one vector multiplet. Concluding Sect. 5 contains the summary and the outlook. Our spinor conventions are given in Appendix A. Some technical details related to the realizations of the $su(2)$--algebra in terms of phase space functions are gathered in Appendix B.

\vspace{0.5cm}

\noindent
{\bf 2. Extending $su(2)$ to $su(1,1|2)$}\\

\noindent
Let us consider a phase space parametrized by the canonical pairs $(\theta^A, p_{\theta A})$, $A=1,\dots,n$, which obey the conventional Poisson brackets\footnote{Here and in what follows the vanishing Poisson brackets are omitted.}
\be
\{\theta^A,p_{\theta B} \}={\delta^A}_B
\ee
and assume that on such a phase space one can construct the functions $J_a=J_a (\theta,p_\theta)$, $a=1,2,3$, which obey the structure relations of $su(2)$
\be\label{su(2)}
\{J_a,J_b \}=\e_{abc} J_c,
\ee
where $\e_{abc}$ is the Levi-Civita totally antisymmetric symbol with $\epsilon_{123}=1$.
Then one can construct an $su(2)$--invariant dynamical system by identifying its Hamiltonian with the Casimir element of the algebra
\be\label{H1}
H_0=\frac 12 J_a J_a.
\ee

As is well known, $su(2)$ is the $R$--symmetry subalgebra of the superalgebra $su(1,1|2)$. It is then natural to expect that each dynamical system like (\ref{su(2)}), (\ref{H1}) can be extended to accommodate the full $N=4$ superconformal symmetry. In order to demonstrate this, let us extend the phase space $(\theta^A, p_{\theta A})$ by an extra bosonic pair $(x,p)$, a fermionic $SU(2)$--spinor variable $\psi_\alpha$, $\alpha=1,2$, its complex conjugate ${(\psi_\alpha)}^{*}=\bar\p^\a$, and impose the canonical brackets
\be\label{cr}
\{x,p\}=1\ , \qquad \{ \p_\a, \bar\p^\b \}=-i{\d_\a}^\b.
\ee
It is assumed that the Poisson brackets of the new variables with the pair $(\theta^A, p_{\theta A})$ vanish. Note that in the subsequent sections $(\theta^A, p_{\theta A})$ and $(x,p)$ are going to be the angular and the radial degrees of freedom of a superconformal system.

The canonical relations (\ref{su(2)}) and (\ref{cr}) are all one needs to use in order to verify that the following functions\footnote{
In Eq. (\ref{rep}) $\s_a$ designate the Pauli matrices. Our spinor notations are gathered in Appendix A.}:
\begin{align}
&
H=\frac{p^2}{2}+\frac{2}{x^2} J_a J_a-\frac{2}{x^2} (\bar\p \s_a \p) J_a +\frac{1}{4x^2} \p^2 \bar\p^2, && D=tH-\frac 12 x p,
\nonumber\\[2pt]
&
K=t^2 H-t x p +\frac 12 x^2, && \mathcal{J}_a=J_a+\frac 12 (\bar\p \s_a \p),
\nonumber\\[2pt]
&
Q_\a=p \p_\a+\frac{2i}{x} {(\s_a \p)}_\a J_a +\frac{i}{2x} \bar\p_\a \p^2\ , && S_\a=x \p_\a -t Q_\a,
\nonumber
\end{align}
\begin{align}\label{rep}
&
\bar Q^\a =p \bar\p^\a-\frac{2i}{x} {(\bar\p \s_a)}^\a J_a +\frac{i}{2x} \p^\a \bar\p^2, &&
\bar S^\a=x \bar\p^\a -t \bar Q^\a,
\end{align}
obey the structure relations of $su(1,1|2)$
\begin{align}\label{algebra}
&
\{ H,D \}=H, && \{ H,K \}=2D,
\nonumber\\[2pt]
&
\{D,K\}=K, && \{ \mathcal{J}_a,\mathcal{J}_b \}=\epsilon_{abc} \mathcal{J}_c,
\nonumber\\[2pt]
&
\{ Q_\a, \bar Q^\b \}=-2 i H {\d_\a}^\b, &&
\{ Q_\a, \bar S^\b \}=2{{(\s_a)}_\a}^\b \mathcal{J}_a+2iD {\d_\a}^\b,
\nonumber\\[2pt]
&
\{ S_\a, \bar S^\b \}=-2i K {\d_\a}^\b, &&
\{ \bar Q^\a, S_\b \}=-2{{(\s_a)}_\b}^\a \mathcal{J}_a+2iD {\d_\b}^\a,
\nonumber\\[2pt]
& \{ D,Q_\a\} = -\frac{1}{2} Q_\a, && \{ D,S_\a\} =\frac{1}{2} S_\a,
\nonumber\\[2pt]
&
\{ K,Q_\a \} =S_\a, && \{ H,S_\a \}=-Q_\a,
\nonumber\\[2pt]
&
\{ \mathcal{J}_a,Q_\a\} =\frac{i}{2} {{(\s_a)}_\a}^\b Q_\b, && \{ \mathcal{J}_a,S_\a\} =\frac{i}{2} {{(\s_a)}_\a}^\b S_\b,
\nonumber\\[2pt]
& \{ D,\bar Q^\a \} =-\frac{1}{2} \bar Q^\a, && \{ D,\bar S^\a\} =\frac{1}{2} \bar S^\a,
\nonumber\\[2pt]
& \{K,\bar Q^\a\} =\bar S^\a, && \{ H,\bar S^\a\} =-\bar Q^\a,
\nonumber\\[2pt]
&
\{\mathcal{J}_a,\bar Q^\a\} =-\frac{i}{2} \bar Q^\b {{(\s_a)}_\b}^\a, && \{ \mathcal{J}_a,\bar S^\a\} =-\frac{i}{2}
\bar S^\b {{(\s_a)}_\b}^\a.
\end{align}
In dynamical realizations $H$ is interpreted as the Hamiltonian. $D$, $K$, $\mathcal{J}_a$ are treated as the generators of dilatations, special conformal transformations, and $su(2)$--transformations, respectively. $Q_\a$ is the supersymmetry generator and $S_\a$ is the generator of superconformal transformations.
In verifying the structure relations (\ref{algebra}), the properties of the Pauli matrices and the spinor identities exposed in Appendix A prove to be helpful.

Note that the dynamical realization (\ref{rep}) of the superalgebra $su(1,1|2)$ involves $(n+1)$ bosonic degrees of freedom and $4$ real fermions. Thus, except for $n=3$,  (\ref{rep}) is an on--shell formulation.
Setting the fermions to zero and $x$ to a constant value, one recovers the original $su(2)$--invariant dynamical system governed by (\ref{H1}).

\vspace{0.5cm}

\noindent
{\bf 3. Examples}\\

\noindent
A trivial realization of the scheme above is provided by
\be\label{empty}
J_a=0,
\ee
which means that the angular sector is empty. In this case there is only one dynamical boson $x$ and four dynamical fermions $(\psi_\alpha,\bar\psi^\alpha)$.
Eq. (\ref{rep}) corresponds to a particular (on-shell) dynamical realization of the superalgebra $su(1,1|2)$ in terms of the $(1,4,3)$--supermultiplet.

A more interesting example is obtained by restricting the conventional angular momentum $J_a=\epsilon_{abc} x_b p_c$ written in Cartesian coordinates to the surface of a two--dimensional sphere of the unit radius
\bea\label{S2}
&&
J_1=-p_\Phi \cot \Theta \cos \Phi  -p_\Theta \sin \Phi, \quad J_2=-p_\Phi \cot \Theta \sin \Phi +p_\Theta \cos \Phi, \quad J_3=p_\Phi,
\eea
where $(\Theta,p_\Theta)$ and $(\Phi,p_\Phi)$ form the canonical pairs.
There are three bosonic dynamical degrees of freedom $(x,\Theta,\Phi)$ and four fermions $(\psi_\alpha,{\bar\psi}^\alpha)$ which all together provide an on--shell dynamical realization
of the superalgebra $su(1,1|2)$ in terms of the $(3,4,1)$--supermultiplet.

As is known, (\ref{S2}) can be deformed to include a contribution which is physically interpreted as due to the magnetic monopole field. It is conventionally described by the shift
\be\label{shift}
p_a \quad \rightarrow \quad p_a+A_a (\Theta,\Phi),
\ee
where the label $a$ takes two values $(\Theta,\Phi)$, and $A_a (\Theta,\Phi)$ is the gauge field potential. Note that the pure gauge vector potential $A_a (\Theta,\Phi)=\partial_a \epsilon(\Theta,\Phi)$, where $\epsilon(\Theta,\Phi)$ is some scalar function, is usually ignored because it can be removed by a canonical transformation $\Theta'=\Theta$, $p'_\Theta=p_\Theta+\partial_\Theta \epsilon(\Theta,\Phi)$, $\Phi'=\Phi$, $p'_\Phi=p_\Phi+\partial_\Phi \epsilon(\Theta,\Phi)$.

Given the representation (\ref{S2}), the shift (\ref{shift}) yields
\be\label{shift1}
J_a \quad \rightarrow \quad J'_a=J_a+B_a(\Theta,\Phi),
\ee
where the explicit form of the functions $B_a(\Theta,\Phi)$ can be readily derived from $A_a (\Theta,\Phi)$ and Eq. (\ref{S2}). Demanding $J'_a$ to obey the structure relations of $su(2)$, one obtains a system of the first order linear partial differential equations which can be solved in full generality (see Appendix B). The result reads
\bea\label{s2}
&&
J'_1=-p_\Phi \cot \Theta \cos \Phi  -p_\Theta \sin \Phi+e \cos \Phi \sin^{-1}\Theta,
\nonumber\\[2pt]
&&
J'_2=-p_\Phi \cot \Theta \sin \Phi +p_\Theta \cos \Phi +e \sin \Phi \sin^{-1}\Theta,
\nonumber\\[2pt]
&&
J'_3=p_\Phi, \quad J_a J_a=p_\Theta^2+{(p_\Phi-e \cos\Theta)}^2 \sin^{-2}\Theta  +e^2,
\eea
where $e$ is an arbitrary constant related to the magnetic charge which causes the magnetic monopole field described by the vector potential $A_a (\Theta,\Phi)$. The corresponding $N=4$ superconformal mechanics describes another variant of the $N=4$, $d=1$ supermultiplet of the type $(3,4,1)$. It has been studied in detail in Ref. \cite{ikl} (see also a related work \cite{g1}).

The next example is obtained from (\ref{s2}) by applying the oxidation procedure. Note that $e$ in (\ref{s2}) is an arbitrary constant. Let us introduce into the consideration an extra canonical pair $(\Psi,p_\Psi)$ and employ in (\ref{s2}) the substitution
\be
e \quad \rightarrow \quad p_\Psi.
\ee
The oxidation is the converse to the reduction in which momenta associated with cyclic variables are set to be coupling constants. The modified functions
\bea\label{s3}
&&
J_1=-p_\Phi \cot \Theta \cos \Phi  -p_\Theta \sin \Phi+p_\Psi \cos \Phi \sin^{-1}\Theta,
\nonumber\\[2pt]
&&
J_2=-p_\Phi \cot \Theta \sin \Phi +p_\Theta \cos \Phi +p_\Psi \sin \Phi \sin^{-1}\Theta,
\nonumber\\[2pt]
&&
J_3=p_\Phi, \quad J_a J_a=p_\Theta^2+{(p_\Phi-p_\Psi \cos\Theta)}^2 \sin^{-2}\Theta +p_\Psi^2,
\eea
automatically obey the structure relations of $su(2)$ because $\psi$ does not enter explicitly. In Eq. (\ref{s3}) one recognizes the vector fields dual to the conventional left--invariant one--forms defined on the group manifold $SU(2)$. Because there are four dynamical bosons $(x,\Theta,\Phi,\Psi)$ and four dynamical fermions $(\psi_\alpha,\bar\psi^\alpha)$, the corresponding superconformal mechanics describes $N=4$, $d=1$ supermultiplet of the type $(4,4,0)$. This is the only off--shell dynamical realization of $su(1,1|2)$ which arises within our formalism.
An attempt to further extend (\ref{s3}) by including a vector field $B_a (\Theta,\Phi,\Psi)$ like in Eq. (\ref{shift1}) above leads to a pure gauge vector potential.

One might as well try to consider various direct sums constructed from $J_a$ as given in Eqs. (\ref{s2}) and (\ref{s3}) above. It turns out that in this case the metric tensor which enters the Casimir element $J_a J_a$ and controls the kinetic terms for the angular variables is degenerate. This means that some of the degrees of freedom are not described by the conventional second order ordinary differential equations. By this reason, in what follows we disregard such a possibility. The examples above thus seem to exhaust all the realizations of $su(2)$ which can be constructed in terms of bosonic variables.

Our next example is provided by a pair of complex conjugate fermions $\chi_\alpha$,
$\bar\chi^\a={(\chi_\alpha)}^{*}$, $\alpha,\beta=1,2$, which obey the canonical bracket $\{ \chi_\a, \bar\chi^\b \}=-i{\d_\a}^\b$. These can be contracted with the Pauli matrices to yield the following realization of $su(2)$:
\be\label{s4}
\tilde J_a=\frac 12 (\bar\chi \sigma_a \chi).
\ee
While one cannot consistently combine two bosonic realizations of $su(2)$ within a consistent dynamical system with the $su(1,1|2)$--superconformal symmetry, the direct sum of $J_a$ in (\ref{empty}), or (\ref{s2}), or (\ref{s3}) with ${\tilde J}_a$ in (\ref{s4}) is admissible. The resulting (on--shell) models can be interpreted as describing a particular interaction of the $(0,4,4)$--supermultiplet realized on the pair $(\chi_\alpha,\bar\chi^\a)$ with either the $(1,4,3)$--, or $(3,4,1)$--, or $(4,4,0)$--supermultiplet. Below we display the corresponding (on--shell) Lagrangian formulations in the explicit form\footnote{Note that
within the Hamiltonian formalism the canonical bracket $\{ \p_\a, \bar\p^\b \}=-i{\d_\a}^\b$ is conventionally understood as
the Dirac bracket $
{\{A,B \}}_D=\{A,B \}-i\{A,\lambda^\a \}\{\bar\lambda_\a,B \}-i\{A,\bar\lambda_\a \}\{\lambda^\a,B \}$
associated with the fermionic second class constraints
$\lambda^\a={p_\p}^\a-\frac i2 \bar\p^\a=0$ and $\bar\lambda_\a=p_{\bar\p \a}-\frac i2 \p_\a=0$. Here
$({p_\p}^\a,p_{\bar\p \a})$ stand for the momenta canonically conjugate to the variables
$(\p_\a,\bar\p^\a)$, respectively. Choosing the right derivative for the fermionic degrees of freedom, the action functional, which reproduces the Dirac bracket for the fermionic pair, reads
$S=\int dt \left(\frac i2 \bar\p^\a {\dot\p}_\a-\frac i2 {\dot{\bar\p}} {}^\a \p_\a\right)$. Similar consideration applies to the fermionic pair $(\chi_\alpha,\bar\chi^\alpha)$.
}.

Combining $\tilde J_a$ in (\ref{s4}) with $J_a$ in (\ref{empty}) and considering the Legendre transform of the Hamiltonian in (\ref{rep}) with respect to the bosonic momenta, one finds the
(on--shell) Lagrangian describing an interaction of the $(0,4,4)$--, and $(1,4,3)$--supermultiplets
\bea
&&
S=\int dt \left(\frac 12 {\dot x}^2 +\frac i2 \bar\p^\a {\dot\p}_\a-\frac i2 {\dot{\bar\p}} {}^\a \p_\a+\frac i2 \bar\chi^\a {\dot\chi}_\a-\frac i2 {\dot{\bar\chi}} {}^\a \chi_\a-\frac{1}{4 x^2} \psi^2 \bar\psi^2+\frac{3}{4 x^2} \chi^2 \bar\chi^2
\right.
\nonumber\\[2pt]
&&
\left.
\qquad \quad -\frac{1}{x^2} (\bar\psi\psi)(\bar\chi\chi)-\frac{2}{x^2} (\bar\psi \chi) (\bar\chi \psi)\right).
\eea

In a similar fashion one can build the (on--shell) Lagrangian describing a particular coupling of the $(0,4,4)$--, and $(3,4,1)$--supermultiplets
\bea\label{accon}
&&
S=\int dt \left(\frac 12 {\dot x}^2 +\frac i2 \bar\p^\a {\dot\p}_\a-\frac i2 {\dot{\bar\p}} {}^\a \p_\a+\frac i2 \bar\chi^\a {\dot\chi}_\a-\frac i2 {\dot{\bar\chi}} {}^\a \chi_\a+\frac 18 x^2 \left({\dot\Theta}^2
+{\dot\Phi}^2 \sin^2 \Theta \right)-\frac{2 e^2}{x^2}
\right.
\nonumber\\[2pt]
&&
\left.
\quad \quad
+ e  \dot\Phi \cos{\Theta}
+\frac 12 (\bar\p \s_a \p-\bar\chi \s_a \chi) \mathcal{L}_a-\frac {1}{x^2} \p^2 \bar\p^2-\frac{1}{2 x^2} {[(\bar\p \s_a \p-\bar\chi \s_a \chi) \lambda_a]}^2
 \right),
\eea
where
\bea\label{LL}
&&
\mathcal{L}_1=-\dot\Theta \sin{\Phi} -\dot\Phi \sin{\Theta} \cos{\Theta} \cos{\Phi}+\frac{4 e}{x^2} \sin{\Theta} \cos{\Phi},
\nonumber\\[2pt]
&&
\mathcal{L}_2=\dot\Theta \cos{\Phi} -\dot\Phi \sin{\Theta} \cos{\Theta} \sin{\Phi}+\frac{4 e}{x^2} \sin{\Theta} \sin{\Phi},
\nonumber\\[2pt]
&&
\mathcal{L}_3=\dot\Phi \sin^2 {\Theta}+\frac{4 e}{x^2} \cos{\Theta}
\eea
and $\lambda_a$ is the vector parameterizing a point on the unit sphere
\be
\lambda_a=(\cos{\Phi} \sin{\Theta}, \sin{\Phi} \sin{\Theta},\cos{\Theta}).
\ee

For an interaction of the $N=4$, $d=1$ supermultiplets of the type $(0,4,4)$ and $(4,4,0)$ the scheme yields
\bea\label{accon}
&&
S=\int dt \left(\frac 12 {\dot x}^2 +\frac i2 \bar\p^\a {\dot\p}_\a-\frac i2 {\dot{\bar\p}} {}^\a \p_\a+\frac i2 \bar\chi^\a {\dot\chi}_\a-\frac i2 {\dot{\bar\chi}} {}^\a \chi_\a+
\right.
\nonumber\\[2pt]
&&
\left.
\quad
+\frac 18 x^2 \left({\dot\Theta}^2
+{\dot\Phi}^2 \sin^2 \Theta +{(\dot\Psi+\dot\Phi \cos{\Theta} )}^2 \right)
+\frac 12 (\bar\p \s_a \p-\bar\chi \s_a \chi) \mathcal{L}_a-\frac {1}{x^2} \p^2 \bar\p^2
 \right),
\eea
where we abbreviated
\bea\label{L}
&&
\mathcal{L}_1=-\dot\Theta \sin{\Phi} +\dot\Psi \sin{\Theta} \cos{\Phi},
\qquad
\mathcal{L}_2=\dot\Theta \cos{\Phi} +\dot\Psi \sin{\Theta} \sin{\Phi},
\nonumber\\[2pt]
&&
\mathcal{L}_3=\dot\Phi +\dot\Psi \cos{\Theta}.
\eea
In the latter case,
when shuffling between the Lagrangian and Hamiltonian formulations, it proves helpful to use the identity
\be
J_a=\frac{x^2}{4} \mathcal{L}_a-\frac 12 (\bar\chi \sigma_a \chi-\bar\psi \sigma_a \psi),
\ee
which relates $J_a$ in (\ref{s3}) and $\mathcal{L}_a$ in (\ref{L}).

Note that one can readily generalize the analysis above to include $N$ copies of the $(0,4,4)$--super\-multiplet. It suffices to introduce into the consideration $N$ canonically conjugate fermionic pairs $\chi_{A \alpha}$,
$\bar\chi_A^\a={(\chi_{A \alpha})}^{*}$, $\{ \chi_{A \a}, \bar\chi_B^\b \}=-i{\d_\a}^\b \delta_{AB}$, $A,B=1,\dots,N$, and to modify ${\tilde J}_a$ in (\ref{s4}) in the evident way
\be
\tilde J_a=\frac 12 \sum_{A=1}^N (\bar\chi_A \sigma_a \chi_A).
\ee
The resulting models will describe a particular interaction of $N$ copies of the $(0,4,4)$--supermultiplet with each other and with one supermultiplet of the type $(1,4,3)$, or $(3,4,1)$, or $(4,4,0)$.

\vspace{0.5cm}

\noindent
{\bf 4. A link to near horizon black hole geometries}\\

As is known since the work in \cite{BIK3,ikn,bgik}, superconformal mechanics can be written in another coordinate basis, conventionally called the AdS basis, which provides an interesting link to massive superparticles propagating near the extreme black hole horizons. So far such a relation has been established for the $N=4$, $d=1$ superconformal mechanics based on the supermultiplet of the type $(3,4,1)$ and the massive $N=4$ superparticle moving near the horizon of an extreme Reissner-Nordstr\"om black hole carrying both the electric and magnetic charges \cite{bgik,g1}. In this section, we generalize the previous studies and provide the universal formulae valid also for the $N=4$, $d=1$ supermultiplet of the type $(4,4,0)$. In the latter case we also identify the corresponding curved background. For notational simplicity below we denote coordinates in the AdS and conformal bases by the same letters.

Consider a phase space which has the same structure as described in the beginning of Sect. 2 and a dynamical system governed by the Hamiltonian:
\bea\label{h5}
&&
H=\frac{x}{M^2}\left(\sqrt{b^2+{(x p)}^2 +J_a J_a} +b
\right)
\nonumber\\[2pt]
&&
\qquad
-\frac{x}{M^2}
((\bar\psi \sigma_a \psi) J_a-\frac 18 \psi^2 \bar\psi^2) {\left(\sqrt{b^2+{(x p)}^2 +J_a J_a}-b\right)}^{-1},
\eea
where $b$ and $M$ are real constants. Making use of the canonical relations (\ref{su(2)}) and (\ref{cr}), the properties of the Pauli matrices and the spinor algebra exposed in the Appendix A, one can verify that the following functions:
\bea\label{h6}
&&
D=tH+x p, \qquad
K=t^2 H+2t x p+\frac{M^2}{x} \left(\sqrt{b^2+{(x p)}^2 +J_a J_a}-b\right),
\nonumber\\[2pt]
&&
S_\a=\p_\a { \left(\frac{2M^2}{x} \left(\sqrt{b^2+{(x p)}^2 +J_a J_a} -b \right)\right)}^{\frac 12}
-t Q_\a,
\nonumber\\[2pt]
&&
{\bar S}^\a={\bar\p}^\a { \left(\frac{2M^2}{x} \left(\sqrt{b^2+{(x p)}^2 +J_a J_a} -b \right)\right)}^{\frac 12}
-t {\bar Q}^\a,
\nonumber\\[2pt]
&&
Q_\a=-\frac{2\left((x p) \p_\a-i{(\s_a \p)}_\a J_a -\frac i4 \bar\p_\a \p^2 \right) }
{{ \left(\frac{2M^2}{x} \left(\sqrt{b^2+{(x p)}^2 +J_a J_a} -b \right)\right)}^{\frac 12}}\ ,
\nonumber\\[2pt]
&&
{\bar Q}^\a=-\frac{2\left((x p) {\bar\p}^\a+i{(\bar\p \s_a)}^\a J_a -\frac i4 \p^\a {\bar\p}^2 \right) }
{{ \left(\frac{2M^2}{x} \left(\sqrt{b^2+{(x p)}^2 +J_a J_a} -b \right)\right)}^{\frac 12}}\ ,
\nonumber\
\eea
\bea
&&
\mathcal{J}_a=J_a+\frac 12 (\bar\p \s_a \p),
\eea
along with the Hamiltonian (\ref{h5})
do obey the structure relations of the superalgebra $su(1,1|2)$.

Note that discarding the fermions $\psi_\alpha$ in (\ref{h5}) one obtains a bosonic system whose structure looks typical for a massive relativistic particle propagating on a curved background. A link between the AdS basis (\ref{h5}), (\ref{h6}) and the conformal basis (\ref{rep}) is provided by the canonical transformation  (for more details see \cite{bgik,g1,gn})
\bea
&&
x'={ \left[\frac{2M^2}{x} \left(\sqrt{b^2+{(x p)}^2 +J_a J_a}-b \right)\right]}^{\frac 12}\ ,
\nonumber\\[2pt]
&&
p'=-2 xp { \left[\frac{2M^2}{x} \left(\sqrt{b^2+{(x p)}^2 +J_a J_a} -b \right)\right]}^{-\frac 12},
\nonumber\\[2pt]
&&
J'_a=J_a, \quad \psi'_\alpha=\psi_\alpha,
\eea
where the prime denotes the coordinates in the conformal basis.

Let us discuss which curved backgrounds there correspond to the bosonic realizations of $su(2)$ given in Sect. 3. The second and third examples yield the model of an $N=4$ superparticle
propagating near the horizon of an extreme Reissner-Nordstr\"om black hole which carries either electric, or both the electric and magnetic charges \cite{bgik,g1}.
This can be seen by redefining $e \rightarrow e p$ in Eq. (\ref{s2}) above, where $p$ is a constant,
and inserting the following values:
\be\label{param}
b=m M=e q, \qquad M=\sqrt{q^2+p^2}
\ee
into the formulae (\ref{h5}), (\ref{h6}).
The parameters $m$ and $e$ are interpreted as the mass of a particle probe and its electric charge, while $M$, $q$, $p$ denote the mass of the black hole, and its electric and magnetic charges, respectively.

A curved background associated with the realization (\ref{s3}) of $su(2)$ has not yet been discussed in the literature. In the remaining part of this section we dwell on this issue.
Because the fermionic degrees of freedom are inessential for identifying the background, in what follows we disregard them.

Taking into account the conventional form of the action functional which describes a massive relativistic particle coupled to a curved background and a vector potential one--form
\be\label{AC}
S=-\int \left(m ds+e A \right),
\ee
where $m$ is the mass and $e$ is the electric charge of a particle, one can readily obtain the metric and the vector potential which lead to the Hamiltonian (\ref{h5}) with $J_a J_a$ given in (\ref{s3})
\bea\label{backfields}
&&
ds^2=g_{mn} dx^m dx^n
={\left(\frac{r}{M}\right)}^2 dt^2-{\left(\frac{M}{r}\right)}^2 dr^2-M^2 d \Omega_3^2, \quad A=A_n dx^n=\frac{r}{M} dt,
\nonumber\\[2pt]
&&
d \Omega_3^2=d \Theta^2+\sin^2{\Theta} d \Phi^2+{(d \Psi+\cos{\Theta} d \Phi)}^2.
\eea
The particle (\ref{AC}) thus propagates on the $AdS_2 \times S^3$ background with the two--form flux.

It is straightforward to verify that (\ref{backfields}) does not solve the vacuum Einstein--Maxwell equations in five dimensions. Yet, one can consistently extend the configuration (\ref{backfields}) by other (matter) fields which all together provide a solution of an extended Einstein--Maxwell system. Because a massive relativistic particle does not couple to those fields, any such background is consistent.

Our first example is provided by the bosonic sector of the $d=5$, $N=2$ supergravity interacting with one vector multiplet \cite{cfgk}~\footnote{We use the mostly minus signature convention for the metric $g_{mn}$ and set $g=\det{g_{mn}}$.}
\bea
&&
S=-\int d^5 x \sqrt{g} \left(R+\frac 12 e^{\frac 23 \varphi } F_{nm} F^{nm}+\frac 12 e^{-\frac 43 \varphi} G_{nm} G^{nm} -\frac 13 \partial_n \varphi \partial^n \varphi
\right.
\nonumber\\[2pt]
&&
\quad \quad
\left.
-\frac{1}{2 \sqrt{2 g}} \epsilon^{mnpqr} F_{mn} F_{pq} B_r\right),
\eea
where $R$ is the scalar curvature, $\varphi$ is a scalar field, and $F_{nm}=\partial_n A_m-\partial_m A_n$, $G_{nm}=\partial_n B_m-\partial_m B_n$. The corresponding equations of motion read
\bea\label{EqM}
&&
R_{mn}-\frac 12 g_{mn} R+e^{\frac 23 \varphi }  (F_{mk} {F_n}^k-\frac 14 g_{mn} F^2)+e^{-\frac 43 \varphi }  (G_{mk} {G_n}^k-\frac 14 g_{mn} G^2)
\nonumber\\[2pt]
&&
-\frac 13 (\partial_m \varphi \partial_n \varphi-\frac 12 g_{mn} \partial_k \varphi \partial^k \varphi)=0, \quad
\nabla_m \left(e^{\frac 23 \varphi } F^{mn}-\frac{1}{\sqrt{2g}} \epsilon^{mnpqr} F_{pq} B_r \right)=0,
\nonumber\\[2pt]
&&
\nabla_m \left(e^{-\frac 43 \varphi } G^{mr} \right)+\frac{1}{4 \sqrt{2 g}} \epsilon^{mnpqr} F_{mn} F_{pq} =0, \qquad \nabla^2 \varphi+\frac 12 e^{\frac 23 \varphi } F^2-e^{-\frac 43 \varphi } G^2=0,
\eea
where we denoted $F^2=F_{nm} F^{nm}$, $G^2=G_{nm} G^{nm}$.
One can readily verify that an extension of (\ref{backfields}) by
\be
B=B_n dx^n=\frac{r}{\sqrt{2} M} dt, \qquad \varphi=0
\ee
does yield a solution of (\ref{EqM}).

Our second example is provided by a variant of the Einstein--Maxwell--dilaton system governed by the action functional
\be
S=-\int d^5 x \sqrt{g} \left(R+U(\varphi) F_{nm} F^{nm}-\frac 12 \partial_n \varphi \partial^n \varphi+V(\varphi) \right),
\ee
where $F_{nm}=\partial_n A_m-\partial_m A_n$ and $U(\varphi)$, $V(\varphi)$ are scalar potentials to be fixed below. In this case the equations of motion read
\bea\label{EEM}
&&
R_{mn}-\frac 12 g_{mn} R+2 U(\varphi) (F_{mk} {F_n}^k-\frac 14 g_{mn} F^2)
-\frac 12 (\partial_m \varphi \partial_n \varphi-\frac 12 g_{mn} [\partial_k \varphi \partial^k \varphi-2V(\varphi)])=0,
\nonumber\\[2pt]
&&
\nabla_m \left(U(\varphi) F^{mn}\right)=0, \qquad \nabla^2 \varphi+U'(\varphi) F^2+V'(\varphi)=0.
\eea
The rightmost relation entering the second line in (\ref{EEM}) and the fact that for $A$ in (\ref{backfields}) the square of the field strength reads $F^2=-\frac{2}{M^2}$ prompts one to choose
the potential $V(\varphi)$ in the form
\be
V(\varphi)=\frac{2}{M^2} U(\varphi)+V_0,
\ee
where $V_0$ is a constant.
Then it is straightforward to verify that a constant value of the scalar field $\varphi$
\be
\varphi=\varphi_0,
\ee
along with (\ref{backfields}) yield a solution of the full system (\ref{EEM}), provided
\be
V_0=-\frac{3}{2 M^2}
\ee
and the potential $U(\varphi)$ is chosen so as to obey the initial condition
\be
U(\varphi_0)=\frac 34.
\ee

Note that, while a background geometry originating from the conformal mechanics based on the $(4,4,0)$--supermultiplet does not seem to be unique, the
$d=5$, $N=2$ supergravity interacting with one vector multiplet seems to be the most natural candidate. In this case the supersymmetry doubling occurs near the horizon \cite{cfgk} and
the number of the supercharges in the $N=4$ mechanics matches the number of the Killing spinors characterizing the background geometry. A possibility to interpret the second example above as a supersymmetric
solution deserves a further investigation.

\vspace{0.5cm}

\noindent
{\bf 5. Conclusion}

\vspace{0.5cm}

To summarize, in this work we reconsidered the issue of constructing an $N=4$ superconformal mechanics in one dimension with a special emphasis put on the role played by the $su(2)$--subgroup. Possible realizations of $su(2)$ in terms of bosonic and fermionic degrees of freedom have been considered. It was demonstrated that arranging the $su(2)$--generators so as to include both bosons and fermions one can construct novel dynamical realizations of the superalgebra
$su(1,1|2)$. They can be interpreted as describing an interaction of the $N=4$, $d=1$ supermultiplet of the type $(0,4,4)$ with one of the supermultiplets of the type $(1,4,3)$, or $(3,4,1)$, or $(4,4,0)$. A relation between
the $N=4$, $d=1$ superconformal mechanics and the massive superparticles propagating near the black hole horizons has been discussed. The background geometry associated with the superconformal mechanics based on the $(4,4,0)$--supermultiplet has been identified with the near horizon limit of the $d=5$, $N=2$ supergravity interacting with one vector multiplet.

There are several directions in which the present work can be extended. First of all, it would be interesting to construct off--shell superfield formulations for the component actions presented in Sect. 3. In this respect it is important to understand whether the coupling of the $(1,4,3)$--, and $(0,4,4)$--supermultiplets constructed in Sect. 3 can be linked to the superfield models presented in \cite{DI1} (see also an earlier related work \cite{DI}).
Note that the field content of the systems presented in Sect. 3 indicates that they might exhibit hidden $N=8$ superconformal symmetry (in this respect see also \cite{DI1}). The latter issue deserves a special investigation.
A possibility to extend the present analysis to the case of $D(2,1;\alpha)$--supergroup is worth studying as well (see a related recent work \cite{Hid}). Our analysis in Sect. 4 did not cover the cases of the $N=4$, $d=1$ supermultiplets of the type $(1,4,3)$ and $(2,4,2)$ because on the curved background side there is no room for the rotation symmetry (the two--dimensional and three--dimensional backgrounds, respectively).
It is of interest to study an $N<4$ superconformal mechanics associated with the $d<4$ near horizon backgrounds.

\vspace{0.5cm}

\noindent{\bf Acknowledgements}\\

\noindent
We thank Alessio Marrani for the useful comments on the material presented in Sect. 4.
This work was supported by the MSE program Nauka under
the project 3.825.2014/K, the TPU grant LRU.FTI.123.2014, and the RFBR-DFG grant No 13-02-91330.

\noindent

\vspace{0.5cm}

\noindent
{\bf Appendix A}

\vspace{0.5cm}
\noindent
Throughout the text $SU(2)$--spinor
indices are raised and lowered with the use of the invariant
antisymmetric matrices
\be
\p^\a=\e^{\a\b}\p_\b, \quad {\bar\p}_\a=\e_{\a\b} {\bar\p}^\b,
\nonumber
\ee
where $\e_{12}=1$, $\e^{12}=-1$. Introducing the notation for the spinor bilinears
\be
\quad \p^2=(\p^\a \p_\a\ ) , \quad
\bar\p^2=(\bar\p_\a \bar\p^\a ), \quad \bar\p \p=(\bar\p^\a \p_\a ),
\nonumber
\ee
one gets
\bea
&&
\p_\a \p_\b=\frac 12 \e_{\a\b} \p^2, \qquad \p_\a \bar\chi_\b-\p_\b \bar\chi_\a=\e_{\a\b} (\bar\chi \p),
\nonumber\\[2pt]
&&
\bar\p^\a \bar\p^\b=\frac 12 \e^{\a\b} \bar\p^2,  \qquad \psi^\alpha {\bar\chi}^\beta-\psi^\beta {\bar\chi}^\alpha=-\epsilon^{\alpha \beta} (\bar\chi\psi).
\nonumber
\eea
The Pauli matrices ${{(\s_a)}_\a}^\b$
are chosen in the standard form
\be
\s_1=\begin{pmatrix}0 & 1\\
1 & 0
\end{pmatrix}\ , \qquad \s_2=\begin{pmatrix}0 & -i\\
i & 0
\end{pmatrix}\ ,\qquad
\s_3=\begin{pmatrix}1 & 0\\
0 & -1
\end{pmatrix}\ ,
\nonumber
\ee
which obey
\bea
&&
{{(\s_a \s_b)}_\a}^\b +{{(\s_b \s_a)}_\a}^\b=2 \d_{ab} {\d_\a}^\b \ , \quad
{{(\s_a \s_b)}_\a}^\b -{{(\s_b \s_a)}_\a}^\b=2i \e_{abc} {{(\s_c)}_\a}^\b \ ,
\nonumber\\[2pt]
&&
{{(\s_a \s_b)}_\a}^\b=\d_{ab} {\d_\a}^\b +i \e_{abc} {{(\s_c)}_\a}^\b \ , \quad
{{(\s_a)}_\a}^\b {{(\s_a)}_\g}^\r=2 {\d_\a}^\r {\d_\g}^\b-{\d_\a}^\b {\d_\g}^\r\ ,
\nonumber\\[2pt]
&&
{{(\s_a)}_\a}^\b \e_{\b\g} ={{(\s_a)}_\g}^\b \e_{\b\a}\ , \quad \e^{\a\b} {{(\s_a)}_\b}^\g=\e^{\g\b} {{(\s_a)}_\b}^\a \ ,
\nonumber
\eea
where $\e_{abc}$ is the totally antisymmetric tensor, $\e_{123}=1$. Throughout the text we denote
$\bar\p \s_a \p=\bar\p^\a {{(\s_a)}_\a}^\b \p_\b$. Our conventions for complex conjugation read
\bea
&&
{(\psi_\alpha)}^{*}=\bar\psi^\alpha, \qquad {(\bar\psi_\alpha)}^{*}=-\psi^\alpha, \qquad
{(\psi^2)}^{*}=\bar\psi^2, \qquad {(\bar\psi \sigma_a \chi)}^{*}=\bar\chi \sigma_a \psi.
\nonumber
\eea

\vspace{0.5cm}

\noindent
{\bf Appendix B}

\vspace{0.5cm}
\noindent
Let us consider the set of functions
\be
\quad J'_a=J_a+B_a(\theta,\vf),
\nonumber
\ee
with $J_a$ given in Eq. (\ref{S2}) above, and require $J'_a$ to obey the structure relations of $su(2)$. This yields the system of the first order linear partial differential equations
\bea\label{constrB}
&&
\partial_\vf  B_2 -\partial_\theta  B_3 \cos{\vf}+\partial_\vf  B_3 \cot{\theta} \sin{\vf}-B_1=0,
\nonumber\\[2pt]
&&
\partial_\vf  B_1 +\partial_\theta  B_3 \sin{\vf}+\partial_\vf  B_3 \cot{\theta} \cos{\vf}+B_2=0,
\nonumber\\[2pt]
&&
\partial_\theta B_1  \cos{\vf} -\partial_\vf  B_1 \cot{\theta} \sin{\vf}  +  \partial_\theta B_2 \sin{\vf} + \partial_\vf B_2 \cot{\theta} \cos{\vf} -B_3=0.
\nonumber
\eea
Multiplying the first equation by $-\cos{\vf}$, the second equation by $\sin{\vf}$, and taking the sum, one gets
\bea\label{1}
&&
\partial_\theta B_3=\partial_\vf (B_2 \cos{\vf}-B_1 \sin{\vf}) \quad \Rightarrow \quad
B_2 \cos{\vf}-B_1 \sin{\vf}=\partial_\theta \epsilon(\theta,\vf), \quad B_3=\partial_\vf \epsilon(\theta,\vf),
\nonumber
\eea
where $\epsilon(\theta,\vf)$ is an arbitrary function.
In a similar fashion, the sum of the first equation, which is multiplied by $\sin{\vf}$, and the second equation, which is multiplied by $\cos{\vf}$, gives
\be\label{2}
\partial_\vf (B_1 \cos{\vf}+B_2 \sin{\vf}+B_3 \cot{\theta})=0 \quad \Rightarrow \quad B_1 \cos{\vf}+B_2 \sin{\vf}+B_3 \cot{\theta}=\lambda(\theta),
\nonumber
\ee
with arbitrary function $\lambda(\theta)$. It is then straightforward to verify that the third equation entering the system above reduces to the ordinary differential equation
\be\label{3}
\lambda'(\theta) +\cot{\theta} \lambda(\theta)=0 \quad \Rightarrow \quad \lambda(\theta)=e \sin^{-1} {\theta},
\nonumber
\ee
where $e$ is an arbitrary constant. At this stage, $B_1$ and $B_2$ can be found by purely algebraic means
\bea
&&
B_1=e \cos{\vf} \sin^{-1}{\theta}-\partial_\vf \epsilon \cot{\theta} \cos{\vf}-\partial_{\theta} \epsilon  \sin{\vf},
\nonumber\\[2pt]
&&
B_2=e \sin{\vf} \sin^{-1}{\theta}-\partial_\vf \epsilon  \cot{\theta} \sin{\vf}+\partial_{\theta} \epsilon \cos{\vf},
\nonumber\\[2pt]
&&
B_3=\partial_\vf \epsilon.
\nonumber
\eea
Comparing these expressions with (\ref{S2}), one concludes that $\epsilon(\theta,\vf)$ generates a pure gauge vector potential which can be discarded, while the rest yields (\ref{s2}).


\begin{thebibliography}{nn}
\bibitem{claus}
P. Claus, M. Derix, R. Kallosh, J. Kumar,
P.K. Townsend, A. Van Proeyen, {\it Black holes and superconformal mechanics},
Phys. Rev. Lett. {\bf 81} (1998) 4553, hep-th/9804177.
\bibitem{gibb}
G.W. Gibbons, P.K. Townsend, {\it Black holes and Calogero models},
Phys. Lett. B {\bf 454} (1999) 187, hep-th/9812034.
\bibitem{W}
N. Wyllard, {\it (Super)conformal many body quantum mechanics with extended supersymmetry}, J. Math. Phys. {\bf 41} (2000) 2826, hep-th/9910160.
\bibitem{Gal}
A. Galajinsky, {\it Remarks on N = 4 superconformal extension of the Calogero model}, Mod. Phys. Lett. A {\bf 18} (2003) 1493, hep-th/0302156.
\bibitem{bgl}
S. Bellucci, A. Galajinsky, E. Latini, {\it New insight into WDVV equation}, Phys. Rev. D {\bf 71} (2005) 044023, hep-th/0411232.
\bibitem{glp}
A. Galajinsky, O. Lechtenfeld, K. Polovnikov, {\it N=4 superconformal Calogero models}, JHEP {\bf 0711} (2007) 008, arXiv:0708.1075.
\bibitem{glp1}
A. Galajinsky, O. Lechtenfeld, K. Polovnikov, {\it N=4 mechanics, WDVV equations and roots}, JHEP {\bf 0903} (2009) 113, arXiv:0802.4386.
\bibitem{bks}
S. Bellucci, S. Krivonos, A. Sutulin, {\it N=4 supersymmetric 3-particles Calogero model}, Nucl. Phys. B {\bf 805} (2008), arXiv:0805.3480.
\bibitem{FIL}
S. Fedoruk, E. Ivanov, O. Lechtenfeld, {\it Supersymmetric Calogero models by gauging}, Phys. Rev. D {\bf 79} (2009) 105015, arXiv:0812.4276.
\bibitem{KLP}
S. Krivonos, O. Lechtenfeld, K. Polovnikov, {\it  	
N=4 superconformal n-particle mechanics via superspace}, Nucl. Phys. B {\bf 817} (2009) 265, arXiv:0812.5062.
\bibitem{IFL1}
S. Fedoruk, E. Ivanov, O. Lechtenfeld, {\it Superconformal mechanics}, J. Phys. A {\bf 45} (2012) 173001, arXiv:1112.1947.
\bibitem{BIK3}
S. Bellucci, E. Ivanov, S. Krivonos, {\it AdS/CFT equivalence transformation}, Phys. Rev. D {\bf 66} (2002) 086001, hep-th/0206126.
\bibitem{ikn}
E. Ivanov, S. Krivonos, J. Niederle, {\it Conformal and superconformal mechanics revisited}, Nucl. Phys. B {\bf 677} (2004) 485, hep-th/0210196.
\bibitem{bgik}
S. Bellucci, A. Galajinsky, E. Ivanov, S. Krivonos, {\it  	
${AdS}_2/{CFT}_1$, canonical transformations and superconformal mechanics}, Phys. Lett. B {\bf 555} (2003) 99, hep-th/0212204
\bibitem{g1}
A. Galajinsky, {\it  	
Particle dynamics on ${AdS}_2 \times {S}^2$ background with two-form flux}, Phys. Rev. D {\bf 78} (2008) 044014, arXiv:0806.1629.
\bibitem{gn}
A. Galajinsky, A. Nersessian, {\it  	
Conformal mechanics inspired by extremal black holes in d=4 }, JHEP {\bf 1111} (2011) 135, arXiv:1108.3394.
\bibitem{G2}
A. Galajinsky, {\it Particle dynamics near extreme Kerr throat and supersymmetry}, JHEP {\bf 1011} (2010) 126, arXiv:1009.2341.
\bibitem{GO}
A. Galajinsky, K. Orekhov, {\it N=2 superparticle near horizon of extreme Kerr-Newman-AdS-dS black hole}, Nucl. Phys. B {\bf 850} (2011) 339, arXiv:1103.1047.
\bibitem{BelK}
S. Bellucci, S. Krivonos, {\it  	
N=2 supersymmetric particle near extreme Kerr throat}, JHEP {\bf 1110} (2011) 014, arXiv:1106.4453.
\bibitem{ikl}
E. Ivanov, S. Krivonos, O. Lechtenfeld, {\it New variant of N=4 superconformal mechanics}, JHEP {\bf 03} (2003) 014, arXiv:hep-th/0212303.
\bibitem{IL2}
E. Ivanov, O. Lechtenfeld, {\it  	
N=4 supersymmetric mechanics in harmonic superspace}, JHEP {\bf 0309} (2003) 073, hep-th/0307111.
\bibitem{IKL4}
E. Ivanov, S. Krivonos, O. Lechtenfeld, {\it
$N=4$, $d = 1$ supermultiplets from nonlinear realizations of $D(2,1;\alpha)$}, Class. Quant. Grav. {\bf 21} (2004) 1031, hep-th/0310299.
\bibitem{BK2}
S. Bellucci, S. Krivonos, {\it  	
Potentials in N=4 superconformal mechanics}, Phys. Rev. D {\bf 80} (2009) 065022, arXiv:0905.4633.
\bibitem{FIL2}
S. Fedoruk, E. Ivanov, O. Lechtenfeld, {\it  	
New $D(2,1;\alpha)$ Mechanics with Spin Variables}, JHEP {\bf 1004} (2010) 129, arXiv:0912.3508.
\bibitem{KL1}
S. Krivonos, O. Lechtenfeld, {\it Many-particle mechanics with $D(2,1;\alpha)$ superconformal symmetry}, JHEP {\bf 1102} (2011) 042, arXiv:1012.4639.
\bibitem{KT}
Z. Kuznetsova, F. Toppan, {\it D-module representations of N=2,4,8 superconformal algebras and their superconformal mechanics},
J. Math. Phys. {\bf 53} (2012) 043513, arXiv:1112.0995.
\bibitem{GG}
K. Govil, M. G\"unaydin, {\it Minimal unitary representation of $D(2,1;\lambda)$ and its $SU(2)$ deformations and $d=1$, $N=4$ superconformal models}, Nucl. Phys. B {\bf 869} (2013) 111, arXiv:1209.0233.
\bibitem{HT}
N.L. Holanda, F. Toppan, {\it Four types of (super)conformal mechanics: D-module reps and invariant actions}, J. Math. Phys. {\bf 55} (2014) 061703, arXiv:1402.7298.
\bibitem{TO}
T. Okazaki, {\it Membrane quantum mechanics}, Nucl. Phys. B {\bf 890} (2015) 400, arXiv:1410.8180.
\bibitem{cfgk}
A.H. Chamseddine, S. Ferrara, G.W. Gibbons, R. Kallosh, {\it  	
Enhancement of supersymmetry near 5-d black hole horizon}, Phys. Rev. D {\bf 55} (1997) 3647, hep-th/9610155.
\bibitem{DI1}
F. Delduc, E. Ivanov, {\it Gauging N=4 supersymmetric mechanics II: (1,4,3) models from the (4,4,0) ones}, Nucl. Phys. B {\bf 770} (2007) 179, hep-th/0611247.
\bibitem{DI}
F. Delduc, E. Ivanov, {\it Gauging N=4 supersymmetric mechanics},  Nucl. Phys. B {\bf 753} (2006) 211, hep-th/0605211.
\bibitem{Hid}
T. Hakobyan, S. Krivonos, O. Lechtenfeld, A. Nersessian, {\it Hidden symmetries of integrable conformal mechanical systems}, Phys. Lett. {\bf A} 374 (2010) 801, arXiv:0908.3290.




\end{thebibliography}
\end{document}